\documentstyle[aps,12pt,epsf]{revtex}
\def\beq{\begin{equation}}
\def\eeq{\end{equation}}
\def\bea{\begin{eqnarray}}
\def\eea{\end{eqnarray}}

\def\vel{\left|}
\def\ver{\right|}
\def\nnb{\nonumber}
\def\ga{\left(}
\def\dr{\right)}

\def\nnb{\nonumber}

\def\ba{\begin{array}}
\def\ea{\end{array}}

\def\bos{\lower 0.5cm\hbox{{\vrule width 0pt height 1.3cm}}}
\def\aaa{\lower 0.cm\hbox{{\vrule width 0pt height .8cm}}}
\def\dol{\lower 0.6cm\hbox{{\vrule width 0pt height .8cm}}}

\newcommand\pubdate{\today}

\newcommand\hepnumber{hep-ph/02xxx}
\def\csumb{
$^a$ Institute of Theoretical Physics, Academia Sinica, \\
      P.O.Box 2735, Beijing 100080, P.R.China \\
$^b$ The Abdus Salam International Centre for Theoretical Physics, \\
         P.O.Box 586, 34014 Trieste, Italy                    \\
$^c$ Institute of High Energy Physics, Chinese Academy of Sciences, \\
P.O.Box 918-4, Beijing 100039, P. R. China                     \\
$^d$ Institut f\"ur Theoretische Physik,
Universit\"at Karlsruhe,   \\ D-76128 Karlsruhe, Germany}

\def\Title#1{\begin{center} {\Large\bf #1 } \end{center}}
\def\Author#1{\begin{center}{ \sc #1} \end{center}}
\def\Address#1{\begin{center}{ \it #1} \end{center}}

\newcommand\pubblock{\rightline{\begin{tabular}{l}
         \pubdate\\ \hepnumber \end{tabular}}}

\begin{document}
\begin{titlepage}
\pubblock

\Title{Rare decay $B\rightarrow X_sl^+l^-$ in a CP spontaneously
broken two Higgs doublet model}

\Author{ Chao-Shang Huang $^a$, Wei Liao $^b$, Qi-shu Yan $^c$ and Shou-hua Zhu$^d$ }
\Address{\csumb}
\vfill

\begin{abstract}
The Higgs boson
mass spectrum and couplings of neutral Higgs bosons to fermions are worked out in a CP 
spontaneously broken
two Higgs doublet model in the large tan$\beta$ case.
 The differential branching ratio, forward-backward asymmetry,
CP asymmetry and  lepton polarization for 
$B\rightarrow X_s l^+ l^-$
 are computed. It is shown that effects of neutral Higgs bosons
are quite significant when $\tan\beta$ is large. Especially, the
CP violating normal polarization $P_N$ 
can be as large 
as several percents.

\end{abstract}

\vfill
PACS number(s): 12.60.-i 12.60.Fr 13.20.-v

\end{titlepage}
\eject \baselineskip=0.3in

\section{Introduction}
The recent results on CP violation in 
$B_d$ - $\bar B_d$ mixing have been reported 
by the BaBar and Belle Collaborations 
\cite{Osaka}, which can be explained in the Standard Model (SM)  within
 theoretical and experimental uncertainties.
As it is well-known, the
direct CP violation measurement, Re($\epsilon'$/$\epsilon$), 
in the Kaon system~\cite{kt} can 
also be accommodated by
the CKM phase in the SM within the theoretical uncertainties.
However, the CKM phase is not enough to explain the matter-antimatter 
asymmetry in the universe and gives the contribution to electric dipole moments (EDMs)
of electron and neutron much smaller than the experimental limits
of EDMs of electron and neutron.
Therefore, one needs new sources of CP violation,
which has been one
of motivations to search new theoretical models beyond the SM.

The minimal extension of the SM is to enlarge the Higgs sector
\cite{new2}. It has been shown that if one adheres to the
natural flavor conservation (NFC) in the Higgs sector, then
a minimum of three Higgs doublets are necessary in order to have
spontaneous CP violation \cite{new3}. However, the constraint
can be evaded if one gives up NFC. If NFC is 
broken, one can obtain a so-called general or model III two-Higgs-doublet model (2HDM) 
in which the CP symmetry is explicitly broken. In this paper, however, we will discuss a 
simpler 2HDM in which the Higgs potential 
is CP invariant and $Z_2$ symmetry softly broken. Comparing with the model I or 
model II 2HDM, the Higgs 
potential of this model has an additional linear term of Re($\phi^+_1\phi_2$) and different
self couplings for the real and image parts of $\phi_1^+\phi_2$ \cite{hz,vend,Georgi}.
In this model (we call it model IV 2HDM hereafter) CP symmetry can be spontaneously broken
\cite{hz,vend,Georgi}. So the model IV is the minimal
among the extensions of the SM that provide a new source of CP violation. 
It should be noted that,
in addition to the above terms, if one adds a linear term of 
Im($\phi^+_1\phi_2$), then
one will obtain a CP softly broken 2HDM~\cite{Georgi}. 

Flavor changing neutral current (FCNC) transitions $B\rightarrow X_s\gamma$ and
$B\rightarrow X_sl^+l^-$ provide testing grounds for the SM at
the loop level and sensitivity to new physics.
Rare decays $B\rightarrow X_sl^+l^-(l=e,\mu)$ have been
extensively investigated in both SM and the beyond
 \cite{gsw,ex1}.
In these processes contributions from exchanging neutral
Higgs bosons (NHB) can be safely neglected because of smallness of $\frac{m_l}{m_W}
(l=e,\mu)$ if tan$\beta$ is smaller than about 25.
The inclusive decay $B\rightarrow X_s\tau^+
\tau^-$ has also been investigated in the SM, the model II 2HDM and
SUSY models with and without including the contributions of NHB 
\cite{new7,new8,Dai,kim,ks,add2,lun,h,Bobeth:2001sq,xiong}, 
in a CP softly broken 2HDM~\cite{hz}, as well as in the technicolor
model with scalars \cite{Xiong:2001cp}.
In this paper we extend to investigate  
$B\rightarrow X_s l^+l^- (l=e, \mu, \tau)$
with emphasis on CP violation effects in model IV.
Although there is
little difference between the CP softly 
and spontaneously broken models
\cite{Georgi}, the mass spectrum and consequently 
some phenomenological effects are different.

The paper is organized as follows. 
In section 2, we describe the details of the model IV
and work out the Higgs mass spectrum and couplings of 
Higgs bosons to fermions. Section 3 is devoted
to the effective Hamiltonian responsible for 
$B\rightarrow X_s l^+l^-$. We calculate  
 Wilson coefficients and give all the leading terms.
In Section 4 the formula for CP violating
observables and lepton polarizations in $B\rightarrow X_s l^+l^-$ are given. We give the 
numerical results in section 5. 
Finally, in section 6 we draw conclusions and discussions.

\section{The CP spontaneously broken 2HDM}
For two complex $y=1$, $SU(2)_w$ doublet scalar fields, $\phi_1$ and
$\phi_2$, the simplest Higgs potential, which is NFC softly broken,
can be written as \cite{Georgi}:
\begin{eqnarray}
V(\phi_1,\phi_2) &=&
\sum_{i=1,2} [m_i^2 \phi_i^+ \phi_i +\lambda_i  (\phi_i^+ \phi_i)^2]
+ m_3^2 Re(\phi_1^+\phi_2)+ m_4^2 Im(\phi_1^+ \phi_2) \nonumber\\
&&+ \lambda_3 [ (\phi_1^+ \phi_1)(\phi_2^+ \phi_2) ]
+ \lambda_4
 [ \mbox{Re}(\phi_1^+ \phi_2)]^2
+ \lambda_5
 [ \mbox{Im}(\phi_1^+ \phi_2) ]^2
 \label{eq2}
\end{eqnarray}
Hermiticity requires that all parameters are real. 
It should be noted that
the potential is CP softly broken due to the presence of the term 
$m_4^2 Im(\phi_1^+ \phi_2)$.
We assume that the minimum of the potential is at
\begin{eqnarray}
<\phi_1>=\left( \begin{array}{c}
0 \\
v_1
\end{array}
\right), \ \ \ \
<\phi_2>=\left( \begin{array}{c}
0 \\
v_2 e^{i\xi}
\end{array}
\right),
\end{eqnarray}
 which breaks $SU(2)\times U(1)$
down to $U(1)_{EM}$ and
simultaneously the CP invariance.
The requirement that the vacuum is at least a stationary point of the potential
results in the following three constraints:
\begin{eqnarray}
sin 2\xi v_1 v_2 (\lambda_4 -\lambda_5)+ sin\xi m^2_3 - cos\xi m_4^2 = 0, \nonumber \\
 v_2 cos\xi [2 C_2 + v_1^2 (\lambda_4-\lambda_5)]+v_1 m_3^2 = 0, \nonumber \\
 sin\xi (v_1^2 C_1 - v_2^2 C_2)- v_1 v_2 m_4^2 = 0,
\label{e1}
\end{eqnarray}
where
\begin{eqnarray}
 C_1=m_1^2+2\lambda_1 v_1^2 + (\lambda_3+\frac{\lambda_4+\lambda_5}{2}) v_2^2,\nonumber \\
 C_2=m_2^2+2\lambda_2 v_2^2 + (\lambda_3+\frac{\lambda_4+\lambda_5}{2}) v_1^2.
\end{eqnarray}
For the CP classically invariant case (model IV), 
$m_4^2$=0, eq.($\ref {e1}$) reduces to
\begin{eqnarray}
m_1^2 &=& -[2 \lambda_1 v_1^2 +(\lambda_3+\lambda_5) v_2^2],
\nonumber \\
m_2^2 &=& -[2 \lambda_2 v_2^2 +(\lambda_3+\lambda_5) v_1^2],
\nonumber \\
m_3^2 &=& - 2 v_1 v_2 (\lambda_4 -\lambda_5)\cos\xi.
\label{e2}
\end{eqnarray}
>From eq. ($\ref {e2}$),
one can see that the necessary condition to have spontaneously
broken CP is $\lambda_4\neq \lambda_5$ and $m_3^2 \neq 0$, i.e.,
the real and image parts of $\phi_1^+ \phi_2$ have different
self-couplings and there exists a linear term of
Re($\phi^+_1\phi_2$) in the potential.

We can write the potential at the stationary point as:
\begin{eqnarray}
V&=&m_1^2 v_1^2 +m_2^2 v_2^2+\lambda_1 v_1^4+
\lambda_2 v_2^4+(\lambda_3+\lambda_5) v_1^2 v_2^2 \nonumber\\
&&+(\lambda_4-\lambda_5) v_1^2 v_2^2 \left[
(\cos\xi-\Delta)^2-\Delta^2 \right],
\end{eqnarray}
with
$$
\Delta=-\frac{m_3^2}{2 v_1 v_2 (\lambda_4-\lambda_5)}.
$$
One can see that 
in order for the spontaneous CP-breaking to occur with
$\sin \xi \ne 0$,
the following inequalities must hold:
$$
\lambda_4-\lambda_5 >0, \\
-1 < \Delta < 1,
$$
and the potential minimum is at $\cos\xi = \Delta$, which is
automatically satisfied due to Eq. (\ref{e2}).

In the following we
will work out the mass spectrum of the Higgs bosons in model IV.
 For charged components, the
mass-squared matrix for negative states is
\begin{eqnarray}
-\lambda_5 \left(
\begin{array}{cc}
v_1^2 & -v_1 v_2 e^{i\xi} \\
-v_1 v_2 e^{-i\xi} & v_2^2
\end{array}
\right),
\end{eqnarray}
Diagonalizing the mass-squared matrix results in one
zero-mass Goldstone state:
\begin{eqnarray}
G^-=e^{i\xi} \sin\beta \phi_2^- +\cos\beta \phi_1^-,
\end{eqnarray}
and one massive charged Higgs boson state:
\begin{eqnarray}
H^- &=& e^{i\xi} \cos\beta \phi_2^- - \sin \beta \phi_1^-,
\\
m_{H^-} &=& |\lambda_5| v^2,
\end{eqnarray}
where $tan\beta = v_2/v_1$ and $v^2=v_1^2+v_2^2$, which is
 determined by $2 m_W^2/g^2$.
Correspondingly we could also get the positive states $G^+$ and $H^+$.

For neutral Higgs components, because CP-conservation is broken, the
mass-squared matrix is $4\times 4$, which can not be simply separated into
two $2\times 2$ matrices as usual.
After rotating the would-be Goldestone boson $(v_1 {\rm Im} \phi_1^0+
 v_2 {\rm Im} \phi_2^0)/v$ away,
the elements of the mass matrix of the three physical neutral Higgs bosons
$\mu_{ij}$, in the basis of
\{${\rm Re} \phi_1^0, {\rm Re} \phi_2^0, (v_2 {\rm Im}\phi_1^0-
v_1 {\rm Im}\phi_2^0 )/v$\}, can be written as
\begin{eqnarray}
\mu_{11}&=&4\lambda_1 v_1^2+(\lambda_4-\lambda_5) v_2^2 c^2_{\xi}
\nonumber \\
\mu_{12}&=& v_1 v_2[2 \lambda_3+\lambda_4 c^2_{\xi}+\lambda_5(1+s^2_{\xi})]
 \nonumber \\
\mu_{13}&=&\frac{1}{2} (\lambda_4-\lambda_5) v_2 v s_{2 \xi}
\nonumber \\
\mu_{22}&=&4\lambda_2 v_2^2+(\lambda_4-\lambda_5) v_1^2 c^2_{\xi}
\nonumber \\
\mu_{23}&=&\frac{1}{2}(\lambda_4-\lambda_5) s_{2 \xi} v_1 v
\nonumber \\
\mu_{33}&=&(\lambda_4-\lambda_5) v^2 s^2_{\xi}
\label{e3}
\end{eqnarray}
where $s,c$ represent $\sin, \cos$. In eq. ($\ref {e3}$),
 the constraints in eq. ($\ref {e2}$) have been used.
In the case of large $\tan\beta$ which is what we are interested in
\footnote{In model IV, the fermions obtain masses in the same way
as in model II 2HDM. The contributions to the $B \rightarrow X_s l^+ l^-$
from exchanging neutral Higgs
bosons are enhanced roughly by a factor of $tg^2\beta$.},
if we neglect all terms proportional to $v_1$, i.e., if the parameters $\lambda_i$'s are
of the same order,
one can get from above mass matrix that one of the Higgs boson
masses is zero, 
which is obviously in conflict with current experiments. Therefore, in stead
 we shall discuss the cases in which there is a hierarchy of order of magnitude between the parameters,
$\lambda_1\gg$ other $\lambda$'s, and other terms
proportional to $v_1$ in eq. $(\ref {e3})$ are negligible. For
simplicity, we define $\bar \lambda =\lambda_4-\lambda_5$ and
$\tilde{\lambda}= 4 \lambda_1 v_1^2$. Diagonalizing the
Higgs boson mass-squared matrix results in
\begin{eqnarray}
\left(
\begin{array}{c}
H^0_1\\
H^0_2\\
H^0_3
\end{array}
\right) =
\sqrt{2} \left(
\begin{array}{ccc}
c_\alpha & s_\alpha & 0\\ -s_\alpha & c_\alpha & 0 \\ 0 & 0 & 1
\end{array}
\right)
\left(
\begin{array}{c}
{\rm Im} \phi_1^0 \\
{\rm Re} \phi_1^0 \\
{\rm Re} \phi_2^0
\end{array}
\right)
\label{eq9}
\end{eqnarray}
with masses
\begin{eqnarray}
m_{H^0_1,H^0_2}^2&=& \frac{1}{2} \left( \mu_{11}+\mu_{33} \mp
\sqrt{ (\mu_{11}-\mu_{33} )^2+4 \mu_{13}^2 } \right)
\label{eqmasssquare}
\end{eqnarray}
and the mixing angle
\begin{eqnarray}
\tan (2 \alpha)&=&\frac{ 2 \mu_{13}}{\mu_{33}-\mu_{11}}.
\end{eqnarray}

In model IV, it is assumed that the fermions obtain masses in the same way
as in model II 2HDM. That is, the up-type quarks get masses from
Yukawa couplings to the Higgs doublet $\phi_2$ 
and down-type quarks and leptons get masses from Yukawa
couplings to the Higgs doublet $\phi_1$.
Then it is straightforward to obtain the couplings of neutral Higgs bosons to fermions
\begin{eqnarray}
H^0_1 \bar f f:\ \ \ \ &&  -\frac{i g m_f}{2 m_w c_{\beta}}
(s_\alpha+ i c_\alpha \gamma_5) \nonumber\\ H^0_2 \bar f f: \ \ \
\
 &&  -\frac{i g m_f}{ 2 m_w c_{\beta}} (c_\alpha- i s_\alpha \gamma_5)
 \label{eq11}
\end{eqnarray}
where $f$ represents down-type quarks and leptons. The coupling of
$H_3^0$ to f is not enhanced by tan$\beta$ and will not be given
here explicitly.
 The couplings of the charged Higgs bosons to fermions are 
the same as  those in the
CP-conservative 2HDM (model II, see Ref. \cite{19}).
This is in contrary with the model III~\cite{new4} in which the
couplings of the charged Higgs to fermions can be quite different
from model II. It is easy to see from Eqs. (\ref{eq11})  that the
contributions come from exchanging NHB is proportional to
$\sqrt{2} G_F s_\alpha c_\alpha m_f^2/\cos^2\beta$, so that the
constraint due to EDM translate into the constraint on $\sin
2\alpha \tan^2\beta$ ($1/\cos\beta \sim \tan\beta$ in the large
$\tan\beta$ limit). According to the analysis in Ref. \cite{new6},
we have the constraint
\begin{eqnarray}
\sqrt{|\sin 2\alpha|}\tan\beta < 50 \label{eqa1}
\end{eqnarray}
from the neutron EDM. And the constraint from the electron EDM is
not stronger than Eq. (\ref{eqa1}). It is obvious from Eq.
(\ref{eqa1}) that there is a constraint on $\alpha$
only if $\tan \beta >  50$.

\section{The effective Hamiltonian for $B \rightarrow X_{s} l^{+} l^{-}$ }
As it is well-known, inclusive decay rates of heavy hadrons can be calculated in heavy quark
effective theory (HQET) \cite{17} and it has been shown that the
leading terms in $1/m_Q$
expansion turn out to be the decay of a free (heavy) quark and corrections stem
from the order $1/m_Q^2$ \cite{18}. In what follows we shall
calculate the leading term. The effective Hamiltonian describing the flavor changing processes
$b\rightarrow s l^+ l^-$ can be defined as
\begin{equation}\label{ham}
H_{eff}=\frac{4G_F}{\sqrt{2}}V_{tb}V^*_{ts}(\sum_{i=1}^{10}C_i(\mu)O_i(\mu)
+\sum_{i=1}^{10}C_{Q_i}(\mu)Q_i(\mu))
\end{equation}
where $O_i(i=1,\cdots ,10)$ is the same as that given in the ref.\cite{gsw}, $Q_i$'s
come from exchanging the neutral Higgs bosons and are defined in Ref.
\cite{Dai}. The explicit expressions of the operators governing $B\rightarrow X_sl^+l^-$ are given
as follows:
\begin{eqnarray}
O_7 &=& (e/16\pi^2) m_b (\bar{s}_{L\alpha} \sigma^{\mu\nu}
b_{R\alpha}) F_{\mu\nu}, \nonumber \\
O_8 &=& (e/16\pi^2) (\bar{s}_{L\alpha} \gamma^{\mu}
b_{L\alpha}) \bar l \gamma_{\mu} l, \nonumber \\
O_9 &=& (e/16\pi^2) (\bar{s}_{L\alpha} \gamma^{\mu}
b_{L\alpha}) \bar l \gamma_{\mu} \gamma_5 l, \nonumber \\
Q_1 &=& (e^2/16\pi^2) (\bar{s}_{L\alpha} b_{R\alpha})
 (\bar l l), \nonumber \\
Q_2 &=& (e^2/16\pi^2) (\bar{s}_{L\alpha} b_{R\alpha})
 (\bar l \gamma_5 l).
\end{eqnarray}

For the large $\tan\beta$ case, we can generally write the couplings
as following:
\begin{eqnarray}
H H^\pm G^\mp &:& \ \ \ \  \pm i g C_{H H^+ G^-}, \nonumber \\ 
H H^\pm W ^\mp &:& \ \ \ \ i g C_{HH^+ W^-},  \nonumber \\  
H \bar b b &:& \ \ \ \ i g m_b \tan\beta (C_b+ \bar C_b \gamma_5), \nonumber \\
H \bar l l &:& \ \ \ \ i g m_l \tan\beta (C_l+ \bar C_l \gamma_5). 
\end{eqnarray}
In model VI, we obtain
\begin{eqnarray}
C_{H_1 H^+ G^-} &=&-\sqrt{2} v e^{i \xi} \left[
 c_\alpha (\lambda_4 s_\xi+i \lambda_5 c_\xi)
+s_\alpha (\lambda_4 c_\xi- i \lambda_5 s_\xi+
\tilde{\lambda} ) \right], \nonumber \\
C_{H_2 H^+ G^-} &=&-\sqrt{2} v e^{i \xi} \left[
 -s_\alpha (\lambda_4 s_\xi+i \lambda_5 c_\xi)
+c_\alpha (\lambda_4 c_\xi- i \lambda_5 s_\xi+
\tilde{\lambda} ) \right], \nonumber \\
C_{H_1 H^+ W^-} &=& -\frac{s_\alpha+i c_\alpha}{2}, \nonumber \\
C_{H_2 H^+ W^-} &=& -\frac{c_\alpha-i s_\alpha}{2}. 
\end{eqnarray}
and $C_b, C_l, \bar{C}_b, \bar{C}_l$ can be extracted from Eq. (\ref{eq11}).

At the renormalization point $\mu=m_W$ the coefficients $C_i$'s in
the effective Hamiltonian have been given in the ref.\cite{gsw}
and $C_{Q_i}$'s are (neglecting the $O(tg\beta)$ term)
\begin{eqnarray}
C_{Q_1}(m_W)&=& \frac{m_bm_{l}tg^2\beta x_t}{sin^2\theta_W }
\{ \frac{1}{m_H^2} \left[ 
-m_W^2 (C_b+\bar C_b) f_1+C_{H H^+ G^-} f_2 -m_W C_{H H^+ W^-}
f_2 \right] C_l- \frac{f_3}{ 4 m_W^2} \}, \nonumber \\
C_{Q_2}(m_W)&=& \frac{m_bm_{l}tg^2\beta x_t}{sin^2\theta_W }
\{ \frac{1}{ m_H^2} \left[ 
-m_W^2 (C_b+\bar C_b) f_1+C_{H H^+ G^-} f_2 -m_W C_{H H^+ W^-}
f_2 \right] \bar C_l+ \frac{f_3}{ 4 m_W^2} \}, \nonumber \\
C_{Q_3}(m_W)&=&\frac{m_be^2}{m_{l}g_s^2}(C_{Q_1}(m_W)+C_{Q_2}(m_W)),
\nonumber
\\
C_{Q_4}(m_W)&=&\frac{m_be^2}{m_{l}g_s^2}(C_{Q_1}(m_W)-C_{Q_2}(m_W)),
\nonumber
\\
C_{Q_i}(m_W)&=&0, ~~~~i=5,\cdots, 10 \label{eq1}
\end{eqnarray}
where
\begin{eqnarray}
f_1&=& \frac{x_t ln x_t}{x_t-1}- \frac{x_{H^\pm} ln x_{H^\pm}-x_t
ln x_t }{x_{H^\pm}-x_t}, \nonumber \\ f_2&=& \frac{x_t ln
x_t}{(x_t-1)(x_{H^\pm}-x_t)}- \frac{x_{H^\pm} ln x_{H^\pm}
}{(x_{H^\pm}-x_t)(x_{H^\pm}-1)}, 
\nonumber \\ f_3&=&
\frac{1}{x_{H^\pm}-x_t}(\frac{ln x_t}{x_t-1}-\frac{ ln
x_{H^\pm}}{x_{H^\pm}-1})
\end{eqnarray} with $ x_i=m_i^2/m_w^2$. 
It would be instructive to note that in addition to the diagrams of exchanging neutral
Higgs bosons, the box diagram with a charged Higgs and a W in the loop also gives a
leading contribution proportional to $\tan^2\beta$~\cite{ln,hlyz}.

Neglecting the strange quark mass, the effective Hamiltonian (\ref{ham}) leads
to the following matrix element for $b\rightarrow sl^+l^-$
\begin{eqnarray}\label{matrix}\nonumber
M&=&\frac{G_F\alpha}{\sqrt{2}\pi}V_{tb}V^*_{ts}[C^{eff}_8\bar{s}_L\gamma_{\mu}
b_L\bar{l}\gamma^{\mu}l+C_9\bar{s}_L\gamma_{\mu}b_L\bar{l}\gamma^{\mu}
\gamma^5 l\\
&+&2C_7m_b\bar{s}_Li\sigma^{\mu\nu}\frac{q^{\nu}}{q^2}b_R\bar{l}\gamma^{\mu}
l+C_{Q_1}\bar{s}_Lb_R\bar{l}l+C_{Q_2}\bar{s}_Lb_R\bar{l}\gamma^5 l],
\end{eqnarray}
where \cite{gsw,new7,10}
\begin{eqnarray}\label{coeff}\nonumber
C^{eff}_8&=&C_8+\{g(\frac{m_c}{m_b},\hat{s})\\
&+&\frac{3}{\alpha^2}k\sum_{V_i= J/ \psi,
\psi^{\prime}, \psi^{\prime\prime}...}
\frac{\pi M_{V_i}\Gamma(V_i\rightarrow l^+l^-)}{M^2_{V_i}-q^2
-iM_{V_i}\Gamma_{V_i}}\}(3C_1+C_2),
\end{eqnarray}
with $\hat{s}=q^2/m_b^2,~~q=(p_{\mu^+}+p_{\mu^-})^2$. In (\ref{coeff})
$g(\frac{m_c}{m_b},\hat{s})$ arises from the one-loop matrix element
of the four-quark
operators and can be found in Refs. \cite{gsw,dba}.
 The second term
in the brace in (\ref{coeff}) estimates the long-distance
contribution from the intermediates, J/$\psi$, $\psi^{\prime}$,
$\psi^{\prime\prime}$ ... \cite{gsw,10}. For l=$\tau$, the lowest
resonance J/$\psi$ in the $c\bar c$ system does not contribute
because the invariant mass square of the lepton pair is $s > 4
m_{\tau}^2$.
 In our numerical calculations, we choose $ k (3C_1+C_2)=-0.875$ \cite{PDG}.

The QCD corrections to coefficients $C_i$ and $C_{Q_i}$ can be incooperated
in the standard way by using the renormalization group equations.
Although the $C_i$ at the scale $\mu=O(m_b)$
have been given in the next-to-leading order
approximation (NLO) without including mixing with $Q_i$~\cite{nlo}, we
use the values of $C_i$ only in the leading order approximation (LO)
since no $C_{Q_i}$ have been calculated in NLO.
The $C_i$ and $C_{Q_i}$ with LO QCD corrections at the scale $\mu=O(m_b)$ have been
given in Ref. \cite{Dai}:
\begin{eqnarray}\label{c7}
C_7(m_b)&=&\eta^{-16/23}\left[ C_7(m_W)-[\frac{58}{135}(\eta^{10/23}-1)
+\frac{29}{189}(\eta^{28/23}-1)]C_2(m_W) \right.
\nonumber \\
&&\left. -0.012C_{Q_3}(m_W)\right],
\label{eq18}
\end{eqnarray}
\begin{eqnarray}
C_8(m_b)&=&C_8(m_W)+\frac{4\pi}{\alpha_s(m_W)}[-\frac{4}{33}(1-\eta^{-11/23})
+\frac{8}{87}(1-\eta^{-29/23})]C_2(m_W),\\
C_9(m_b)&=&C_9(m_W),\\
C_{Q_i}(m_b)&=&\eta^{-\gamma_Q/\beta_0}C_{Q_i}(m_W),~~i=1,2,
\end{eqnarray}
where $\gamma_Q=-4$ \cite{21} is the anomalous dimension of $\bar{s}_Lb_R$,
$\beta_0=11-2 n_f/3$, and $\eta=\alpha_s(m_b)/\alpha_s(m_W)$.

After a straightforward calculation,  we obtain the
invariant dilepton mass distribution \cite{Dai}
\begin{eqnarray}
\frac{{\rm d}\Gamma(B\rightarrow X_s l^{+} l^{-})}{{\rm d}s}
 &=& B(B\rightarrow X_c l {\bar \nu}) \frac{{\alpha}^2}
 {4 \pi^2 f(m_c/m_b)} (1-s)^2(1-\frac{4t^2}{s})^{1/2}
 \frac{|V_{tb}V_{ts}^{*}|^2}{|V_{cb}|^2} D(s) \nonumber \\
 D(s) &=& |C_8^{eff}|^2(1+\frac{2t^2}{s})(1+2s)
      + 4|C_7|^2(1+ \frac{2t^2}{s})(1+\frac{2}{s}) \nonumber \\
    & &  + |C_9|^2 [ ( 1 + 2s) + \frac{2t^2}{s}(1-4s)]
      +12 {\rm Re}(C_7 C_{8}^{eff*})(1+\frac{2t^2}{s}) \nonumber \\
  & & + \frac{3}{2}|C_{Q_1}|^2 (s-4t^2) + \frac{3}{2}|C_{Q_2}|^2s
      + 6{\rm Re}(C_9 C_{Q_2}^{*}) t
\label{eq22}
\end{eqnarray}
where s=$q^2/m_b^2$, t=$m_{l}/m_{b}$,
$B(B\rightarrow X_c l {\bar \nu})$ is the branching ratio of $B\rightarrow X_c l {\bar \nu}$,
$f$ is the phase-space factor and f(x)=$1-8 x^2+8 x^6
-x^8-24 x^4 \ln ~ x$.

We also give the forward-backward asymmetry
\begin{eqnarray}
A(s)=\frac{\int^{1}_{0}dz \frac{d^2\Gamma}{ds dz} -
\int^{0}_{-1}dz \frac{d^2\Gamma}{ds dz}}{
\int^{1}_{0}dz \frac{d^2\Gamma}{ds dz} +
\int^{0}_{-1}dz \frac{d^2\Gamma}{ds dz}}
=-3 \sqrt{\frac{1-4 t^2}{s}}
\frac{E(s)}{D(s)}
\end{eqnarray}
where $z=\cos\theta$ and $\theta$ is the angle between the momentum
of the B-meson and that of $l^+$ in the center of mass frame of the
dileptons $l^+l^-$. Here,
\begin{eqnarray}
E(s)={\rm Re} (C_8^{eff} C_9^* s+2 C_7 C_9^*+ C_8^{eff} C_{Q1}^*
t+ 2 C_7 C_{Q2}^* t). \label{eq26}
\end{eqnarray}
\section{CP violating observables and lepton polarizations in $B \rightarrow X_s l^+ l^-$}
The formulas for CP violating observables and lepton polarizations in $B \rightarrow X_s l^+ l^-$
have been given in our previous paper~\cite{hz}. We give the formula below in order to make the
paper self-contained. 
The CP asymmetry for the $B \rightarrow X_s l^+ l^-$ and
$\overline{ B} \rightarrow \overline{ X}_s l^+ l^-$ is commonly
defined as
\begin{eqnarray}
A_{CP}(s)=\frac{{\rm d}\Gamma/{\rm d}s -  {\rm
d}\overline{\Gamma}/{\rm d}s}{ {\rm d}\Gamma/{\rm d}s +  {\rm
d}\overline{\Gamma}/{\rm d}s}.
\end{eqnarray}
The CP asymmetry in the forward-backward asymmetry
for $B \rightarrow X_s l^+ l^-$ and
$\overline{ B} \rightarrow \overline{X}_s l^+l^-$
is defined as
\begin{eqnarray}
B_{CP}(s)= A(s)- \overline{A} (s)
\end{eqnarray}
It is easy to see from Eq. (\ref{eq22})  that  the CP asymmetry
$A_{CP}$, in general, is very small because the weak phase
difference in $C_7 C_8^{eff}$ arises from the small mixing of
$O_7$ with $Q_3$ (see Eq. (\ref{eq18})). 
In contrast to $A_{CP}$, $B_{CP}$ 
can reach a large value when $\tan\beta$ is large, as can be seen from Eqs.
(\ref{eq26}) and (\ref{eq1}). Therefore, we propose to measure
$B_{CP}$ in order to search for new CP violation sources.

Let us now discuss the lepton polarization effects. We define three
orthogonal unit vectors:
\bea
\vec{e}_L &=& \frac{\vec{p}_1}{\vel \vec{p}_1 \ver}~, \nnb \\
\vec{e}_N&=& \frac{\vec{p}_{s} \times \vec{p}_1}
{\vel \vec{p}_{s} \times \vec{p}_1 \ver}~, \nnb \\
\vec{e}_T &=& \vec{e}_N \times \vec{e}_L~, \nnb
\eea
where $\vec{p}_1$ and $\vec{p}_{s}$ are the three momenta of the
$\ell^-$ lepton
and the $s$ quark, respectively, in the center of mass of the
$\ell^+~\ell^-$ system. The differential decay rate for any given spin
direction $\vec{n}$ of the $\ell^-$ lepton, where $\vec{n}$ is a unit vector
in the $\ell^-$ lepton rest frame, can be written as
\bea
\frac{d \Gamma \ga \vec{n} \dr}{{\rm d} s} =
\frac{1}{2} \ga \frac{d \Gamma}{{\rm d} s} \dr_{\!\!\! 0}
\Big[ 1 + \ga P_L\, \vec{e}_L + P_N\, \vec{e}_N + P_T\, \vec{e}_T \dr \cdot
\vec{n} \Big]~,
\label{eq27}
\eea
where the subscript "0" corresponds to the unpolarized case, and $P_L,~P_T$,
and $P_N$, which correspond to the longitudinal, transverse and normal
projections of the lepton spin, respectively, are functions of $s$.
>From Eq. (\ref{eq27}), one has
\bea
P_i (s) = \frac{ {\displaystyle{\frac{d \Gamma}{d s}
\ga \vec{n}=\vec{e}_i \dr -
\frac{d \Gamma}{ds}\ga \vec{n}=-\vec{e}_i \dr}} }
{ {\displaystyle{\frac{d \Gamma}{ds}\ga \vec{n}=\vec{e}_i \dr +
\frac{d \Gamma}{ds}\ga \vec{n}=-\vec{e}_i \dr}} } ~.
\eea

The calculations for the $P_i$'s (i = $L,~T,~N$) lead to the following
results:
\bea
P_L &=&  (1-\frac{4 t^2}{s})^{1/2} \frac{D_L(s)}{D(s)},
\nonumber \\
P_N&=& \frac{3 \pi}{4 s^{1/2}} (1-\frac{4 t^2}{s})^{1/2}
\frac{D_N(s)}{D(s)},
\nonumber \\
P_T&=& -\frac{3 \pi t}{2 s^{1/2}}
\frac{D_T(s)}{D(s)},
\eea
where
\bea
D_L(s) &=& {\rm Re}\left(
 2 (1+2 s) C_8^{eff} C_9^*+12 C_7 C_9^*- 6 t C_{Q_1} C_9^*-
3 s C_{Q_1} C_{Q_2}^* \right), \nonumber \\
D_N(s) &=&  {\rm Im} \left(
2 s  C_{Q_1} C_7^*+s C_{Q_1} C_8^{eff *}+s C_{Q_2} C_9^*+
4 t C_9 C_7^*+ 2 t s C_8^{eff\ *} C_9 \right),
 \nonumber \\
D_T(s) &=&   {\rm Re}\left( -2 C_7 C_9^*+ 4 C_8^{eff} C_7^*
+\frac{4}{s} |C_7|^2- C_8^{eff} C_9^* \right. \nonumber \\ &&
\left. +s |C_8^{eff}|^2 -\frac{s-4 t^2}{2 t} C_{Q_1} C_9^*
-\frac{s}{t} C_{Q_2} C_7^*-\frac{s}{2 t} C_8^{eff} C_{Q_2}^*
\right). \label{eq30} \eea $P_i$ (i=L, T, N) have been given in
the ref. ~\cite{add2}, where there are some errors in $P_T$ and
they gave only two terms in $D_N$, the numerator of $P_N$. We
remind that $P_N$ is the CP-violating projection of the lepton
spin onto the normal of the decay plane. Because $P_N$ in $B
\rightarrow X_s l^+ l^-$ comes from both the quark and lepton
sectors, purely hadronic and leptonic CP-violating observables,
such as $d_n$ or $d_e$, do not necessarily strongly constrain
$P_N$ \cite{add1}. So it is advantageous to use $P_N$ to
investigate CP violation effects in some extensions of SM
\cite{add3}. In the model IV, as pointed out above, $d_n$ and
$d_e$ constrain $\sqrt{|\sin 2\alpha|}\tan\beta$ and consequently
$P_N$ through $C_{Q_i}$ ($i=1,2$) (see Eq. (\ref{eq30})).

\section{Numerical results}
The following parameters have been used in the numerical
calculations: $$ m_{t,{\rm pole}}=175Gev,~m_{b,{\rm
pole}}=5.0Gev,~m_{c,,{\rm pole}}=1.3Gev, $$ $$ ~m_{\mu}=0.105 Gev,
~m_{\tau}=1.777 Gev, ~\eta=1.67. $$ Without losing generality, we
assume  $0 < \xi < 2 \pi$. For Higgs masses, as an example, we choose
$m_{H^\pm}=250 Gev$ (see discussions below), the lightest neutral
Higgs mass fixed to 100 $Gev$, and the heavier neutral Higgs mass being
500 $Gev$. 
It should be pointed out that the region of $\xi$ will be constrained
due to this specific choice of neutral Higgs boson masses ( see
eq. (\ref{eqmasssquare})), which is the reason why there are
gaps in Fig. 1 and Figs. 4-10.
For l=e, the contributions of neutral Higgs bosons are
negligible due
 to smallness of the electron mass so that results are almost the same as those
 in SM. So we only give numerical results for l=$\mu, \tau$. We shall analyse
 the constraint from  $b \rightarrow s \gamma$ in the first subsection and
 give the numerical results for l=$\tau, \mu$ in the second and third
 subsections respectively.

\subsection{The constraint from $b \rightarrow s \gamma$}

Because the couplings of the charged Higgs to fermions in Model IV
are the same as those in the model II, the constraint on
$\tan\beta$ due to effects arising from the charged Higgs are the
same as those in the model II. The constraint on $tg\beta$ from
$K-\bar{K}$ and $B-\bar{B}$ mixing, $\Gamma(b\rightarrow s\gamma)
$,$\Gamma(b\rightarrow c\tau\bar{\nu}_{\tau})$ and $R_b$ has been
given \cite{15}
\begin{equation}
0.7\le tg\beta\le 0.52(\frac{m_{H^{\pm}}}{1Gev})
\end{equation}
(and the lower limit $m_{H^{\pm}}\ge 200 Gev$ has also been given in the ref.
\cite{15}). In Ref. \cite{chbound}, it is pointed out that lower bound of the
charged Higgs is about 250 GeV if one adopts conservative approach to evaluate the theoretical uncertainty; on the other hand, adding different theoretical errors
in  quadrature leads to $m_{H^\pm} > 370$ GeV. Indeed, these bounds are quite
sensitive to the errors of the theoretical predictions and to the details of
the calculations.

Due to the mixing of $O_7$ with $Q_3$, $C_7(\mu)$ is dependent of $C_{Q_3}$ (see eq. (30) ).
So we have to see if the experimental results of $b \rightarrow s \gamma$ impose a constraint
on our model parameters
(see Ref. \cite{Huang:2000tn} for the detail discussion
of the constraint on $C_7$). From the equation \cite{ber,buras}
\begin{equation}
\frac{Br(B \to X_s \gamma)}
     {Br(B \to X_c e \bar{\nu}_e)}=
 \frac{|V_{ts}^* V_{tb}^{}|^2}{|V_{cb}|^2}
\frac{6 \alpha}{\pi f(z)} |C^{\rm eff}_{7}(\mu_b)|^2
\end{equation}
and the experimental results for $b \rightarrow s \gamma$ \cite{bsg}
\begin{equation}
2.0 \times 10^{-4} < Br(b \rightarrow s \gamma) < 4.5 \times 10^{-4},
\end{equation}
we can get the constraint on $|C_7|$. In fig. 1, we show the $C_7$
as a function of $\xi$. One can see from the figure that even for
$\tan\beta=50$, the model can still escape the experimental constraint.

\subsection{$B \rightarrow X_s \tau^+ \tau^-$}

Numerical results for $B \rightarrow X_s \tau^+ \tau^-$ are shown
in Figs. 2-7. From Figs. 2 and 3, we can see that the
contributions of NHBs to the differential branching ratio
$d\Gamma/ds$ and forward-backward asymmetry $A_s$ are significant when $\tan\beta$ is 
50 and the masses of NHBs are in the reasonable
region, which is similar to the case of model II 2HDM without CP
violation \cite{Dai}.

Figs. 4 and 5 are devoted to $B_{CP}$ and $P_N$ as a function of $\xi$.
 From Fig. 4, one can see that $B_{CP}$  can reach about $1.5\%$
for the favorable parameters, and depends strongly on $\xi$.
Fig. 5 shows  that $P_N$
depends also strongly on $\xi$ and can
be as large as $8\%$. It should be noted that experimentally the 
observables after
integrating $s$ are more accessible than those for specific s, therefore  
we present also the integrated-$P_N$  ( the integration range of s 
is 0.6-1 which is apart from the resonance region ) in Fig. 5 and Fig. 8. 
Our numerical results [(b) in Fig. 5] show that the shape of the 
integrated-$P_N$, 
which can also reach several percent,
is similar to that for specific s. For the illumination purpose,
we shall present the results for specific s in most of figures. 

Figs. 6 and 7 show the longitudinal and transverse polarizations
respectively. It is obvious that the contributions of NHBs can
change the polarizations greatly, especially when $\tan\beta$ is large.
The longitudinal  polarization of $B \rightarrow X_s l^+l^-$
has been calculated in SM and several new physics scenarios
\cite{new7}. Switching off the NHB contributions, our results
are in agreement with those in Ref. \cite{new7}.

\subsection{$B \rightarrow X_s \mu^+ \mu^-$}

Because the contributions of NHBs to the differential branching
ratio, forward-backward asymmetry and $B_{CP}$ for the process $B
\rightarrow X_s \mu^+ \mu^-$ are so small if even $tan\beta$ is as 
large as 100, 
which is due to the strong suppression ( $\propto m_l^2/m_w^2$ ),
we do not show
the results here. However, for the lepton polarizations, the
suppression is proportional to $m_l/m_w$, which is not so strong,
and consequently NHBs can
make relatively significant contributions. We 
show the numerical result of $P_N$ and integrated-$P_N$ in Fig. 8,   
$P_L$ and $P_T$ in Figs. 9 and 10.

Fig. 8 shows that $P_N$ 
is sensitive to
$\xi$ and can reach several percent  when $\tan\beta=50$. For $\tan\beta$=10,
$P_N$ is unobservablly small. From Fig. 9-10, one can see that the
contributions of NHBs can change the longitudinal and transverse
polarizations greatly, especially when $\tan\beta$ is large.

\section{Conclusions and discussions}
In summary, we have calculated the differential branching ratio,
back-forward asymmetry, lepton polarizations and some CP violated
observables for $B\rightarrow X_s l^+l^-$ in the model IV 2HDM. As
the main features of the model, NHBs play an important role in
inducing CP violation, in particular, for large $\tan\beta$. We
propose to measure $B_{CP}$ (defined in section IV)
instead of the usual CP asymmetry $A_{CP}$, 
because the former could be observed
for l=$\tau$ if $\tan\beta$ is large enough and
the latter is too small to be observed. The CP violating normal
polarization $P_N$ can reach several percents for l=$\tau$
and $\mu$ when tan$\beta$ is large and Higgs boson masses are in
the reasonable range, which could be observed in the future B
factories with $10^8$- $10^{12}$ B hadrons per year~\cite{bs}. 
It should be noted that 
the results are sensitive
to the mass of the charged Higgs boson. If the charged Higgs boson
is heavy (say $> 400$ GeV), the effects arising from new physics
would disappear. If we take the mass of charged Higgs boson to be 
200 Gev which is
the lowest limit allowed by $B\rightarrow X_s\gamma$, the CP violation effects
will be more significant than those given in the paper. Comparing the results in the paper
with those in the CP softly broken 2HDM, the main difference is the different
$\xi$-dependence. Therefore, it is possible 
to discriminate the model IV from the other 2HDMs by measuring the CP-violated
observables such as $B_{CP}$, $P_N$ if the nature chooses large
$\tan\beta$ and a  light charged Higgs boson.
 Otherwise, it is difficult to discriminate
them.

\section*{Acknowledgements}
This research was supported in part by the National Nature Science
Foundation of China, the Alexander von Humboldt
Foundation.



\newpage

\begin{figure}
\epsfxsize=15 cm \centerline{\epsffile{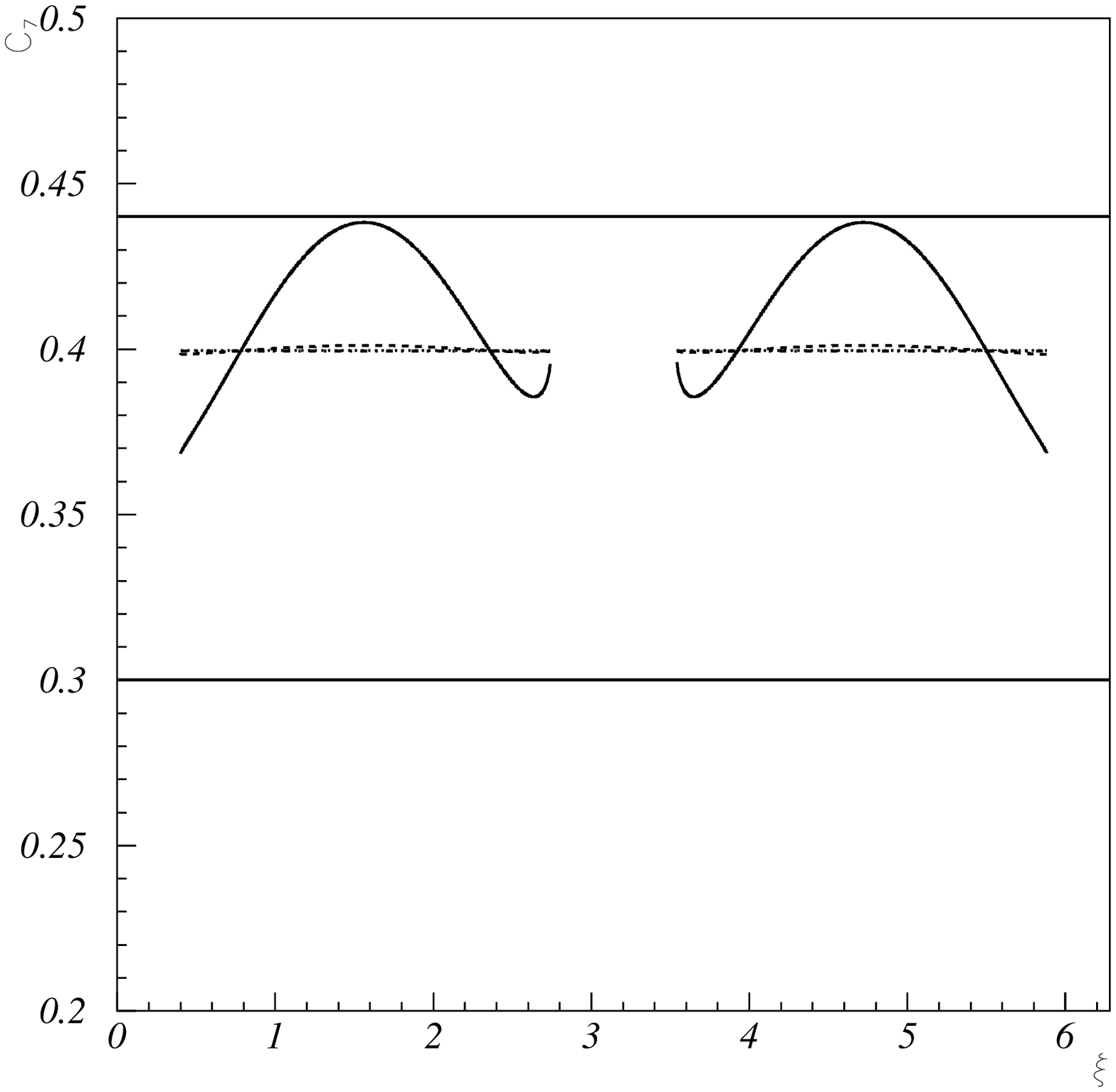}} \caption[]{
$C_7$ as a function of $\xi$ with $m_{H^\pm} = 250 GeV$, and solid
and dashed lines represent $\tan\beta=50$ and $10$, dot-dashed
line represents the case of switching off $C_{Q_i}$ contributions.
The region between two straight solid lines is permitted by the 
 $b \rightarrow s \gamma$ experiment.  }
\end{figure}

\begin{figure}
\epsfxsize=15 cm \centerline{\epsffile{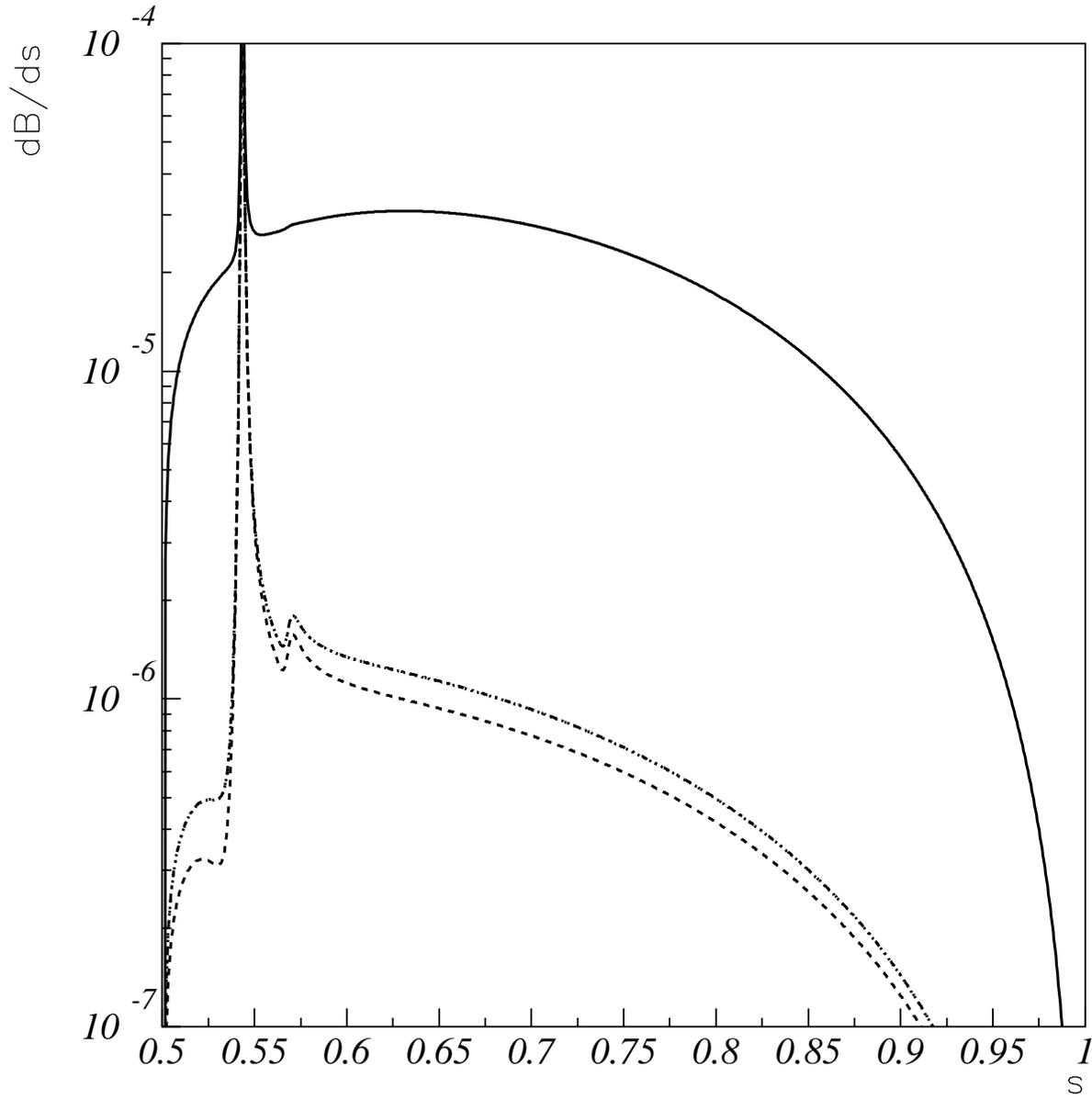}} \caption[]{
Differential branching ratio as function of $s$ for $B \rightarrow
X_s \tau^+\tau^-$, where $\xi=\pi/3$, solid and dashed lines
represent $\tan\beta=50$ and $10$, dot-dashed line represents the
case of switching off $C_{Q_i}$ contributions.} \label{fig1}
\end{figure}

\begin{figure}
\epsfxsize=15 cm
\centerline{\epsffile{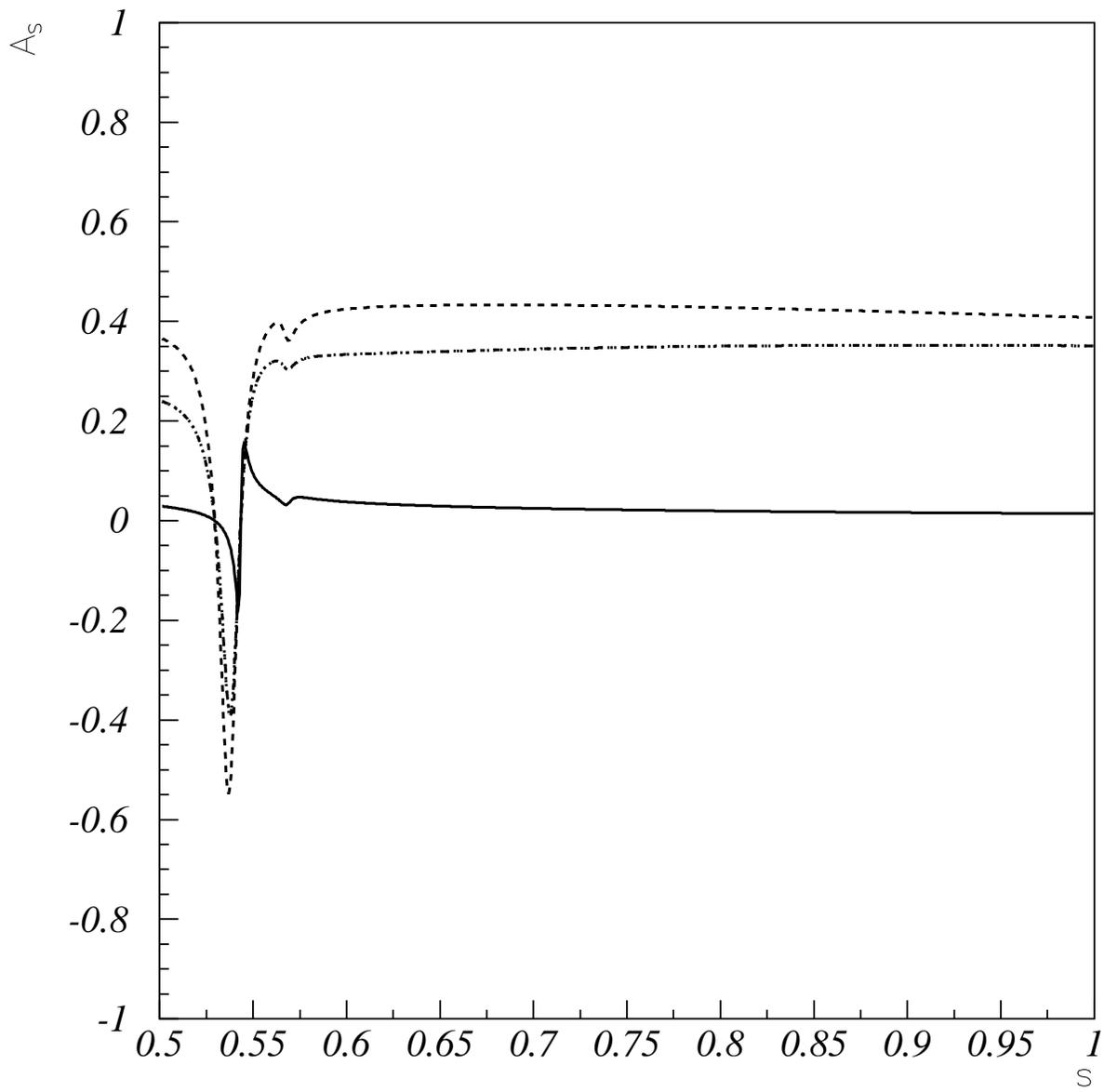}}
\caption[]{
Forward-backward asymmetry as function of $s$, other captions are
same as Fig. \ref{fig1}.}
\end{figure}


\begin{figure}
\epsfxsize=15 cm \centerline{\epsffile{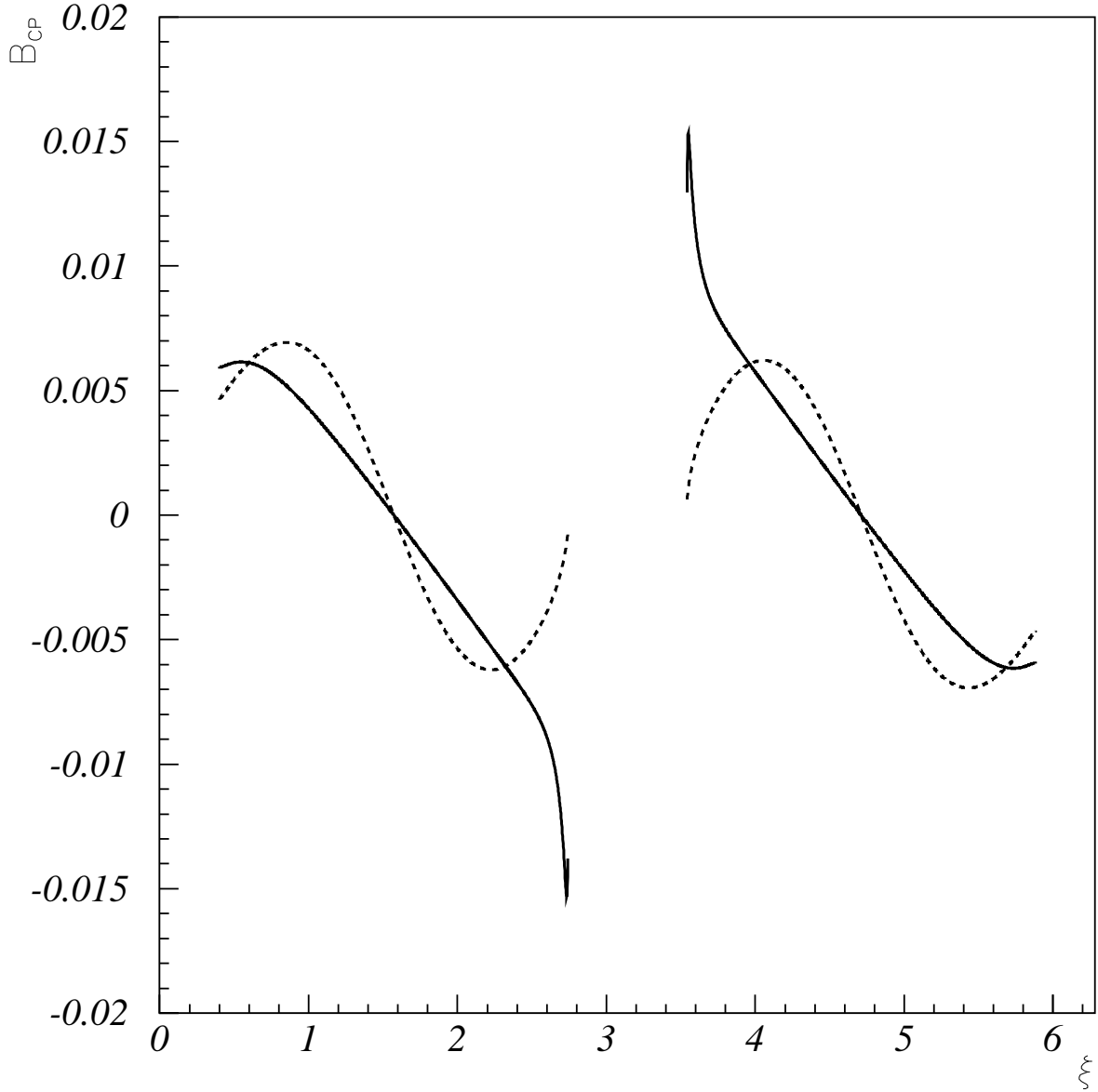}} \caption[]{
$B_{CP}$ as function of $\xi$, for $B \rightarrow X_s
\tau^+\tau^-$, where $s=0.8$,  solid and dashed lines represent
$\tan\beta=50$ and $10$. }
\end{figure}

\begin{figure}
\epsfxsize=15 cm \centerline{\epsffile{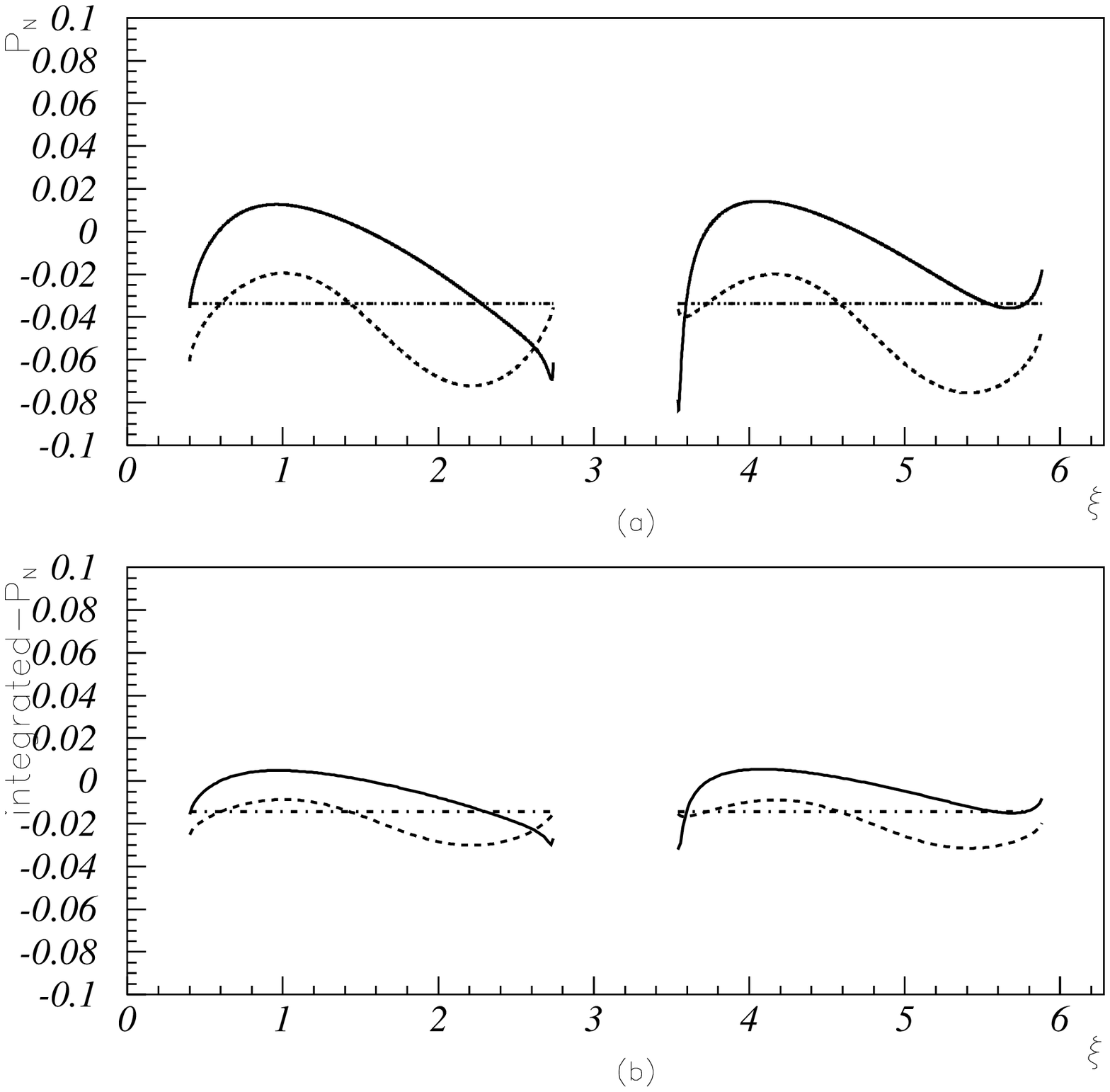}} \caption[]{ $P_N$ (a) and
integrated-$P_N$ (b)
as functions of $\xi$ for $B \rightarrow X_s \tau^+\tau^-$, where
$s=0.8$ for (a),  solid and dashed lines represent $\tan\beta=50$ and
$10$, dot-dashed line represents the case of switching off
$C_{Q_i}$ contributions.} \label{fig2}
\end{figure}

\begin{figure}
\epsfxsize=15 cm
\centerline{\epsffile{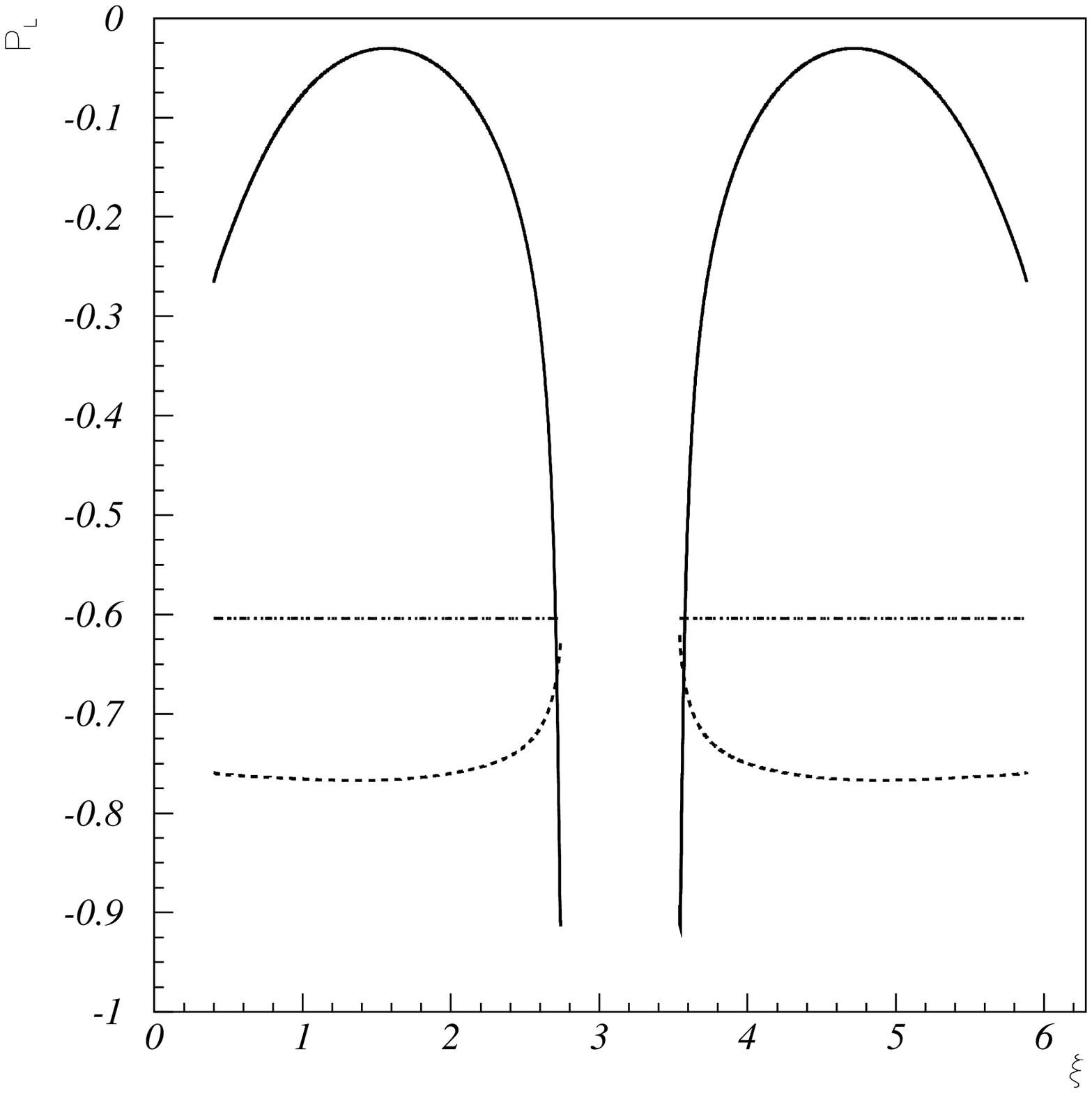}}
\caption[]{
$P_L$ as function of $\xi$,
other captions are same as Fig. \ref{fig2} (a).}
\end{figure}

\begin{figure}
\epsfxsize=15 cm
\centerline{\epsffile{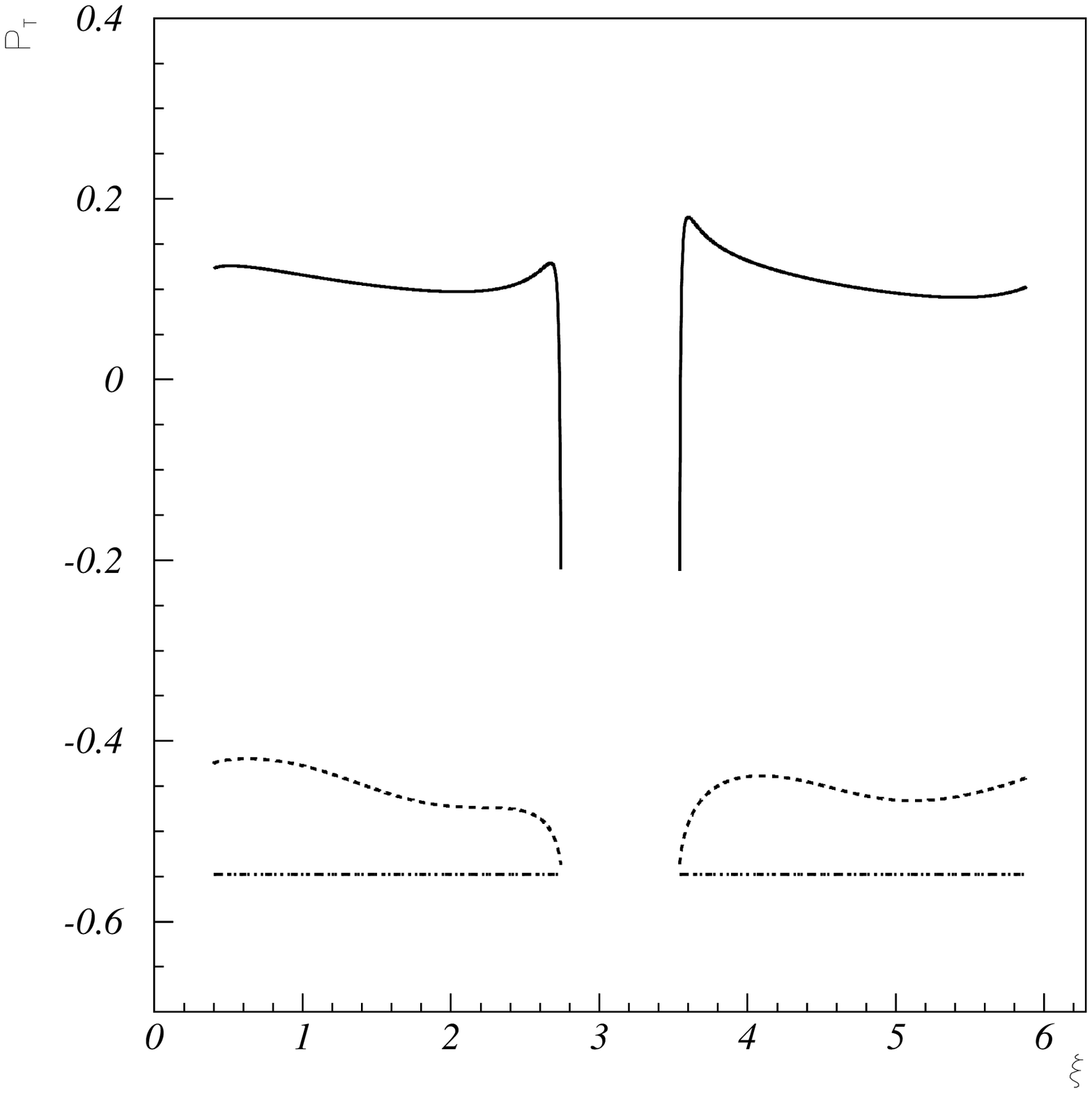}}
\caption[]{
$P_T$ as function of $\xi$,
other captions are same as Fig. \ref{fig2} (a).}
\end{figure}

\begin{figure}
\epsfxsize=15 cm \centerline{\epsffile{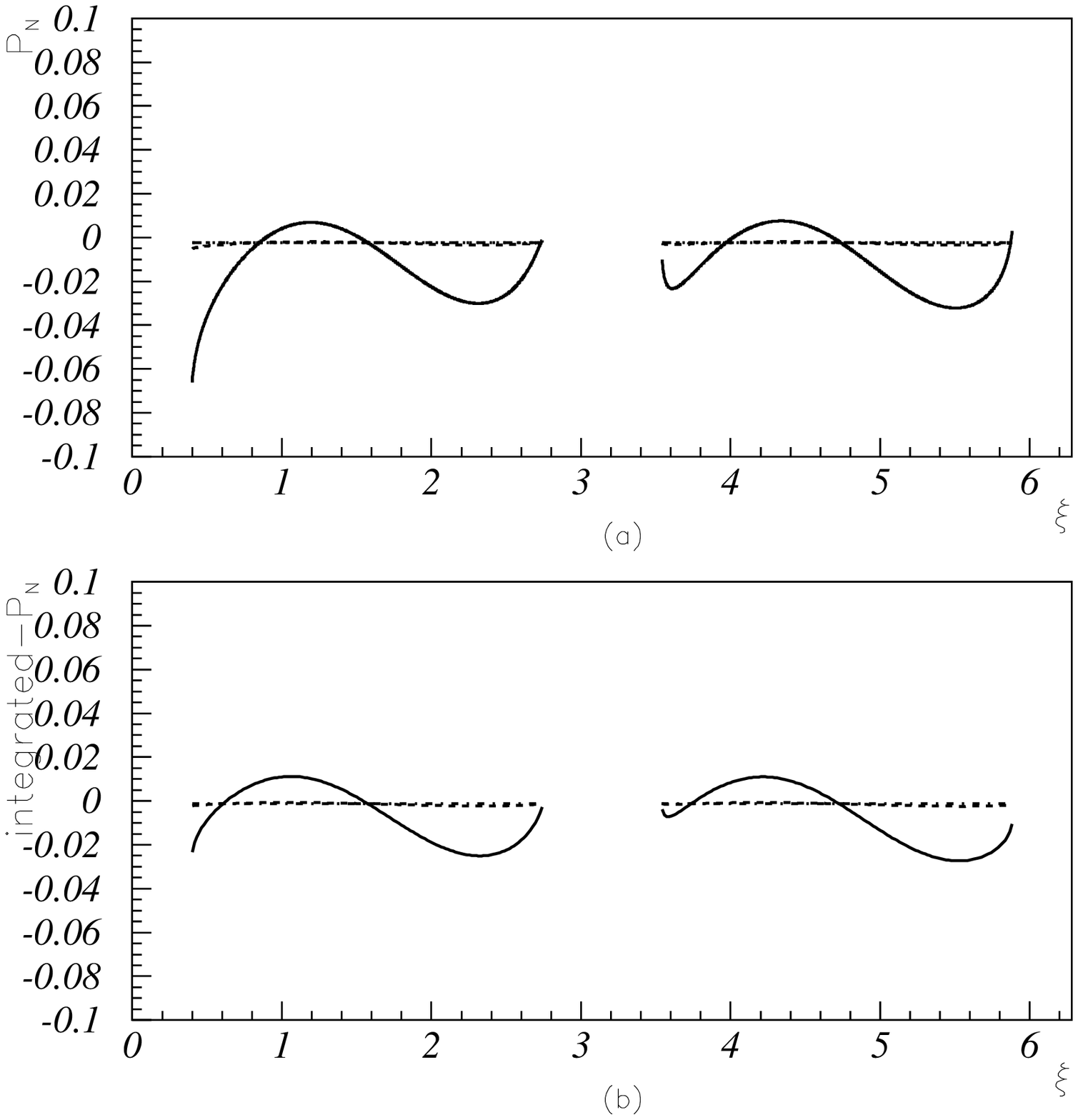}} \caption[]{ $P_N$ (a)
and integrated-$P_N$ (b)
as functions of $\xi$ for $B \rightarrow X_s \mu^+ \mu^-$, where
$s=0.6$ for (a),  solid and dashed lines represent $\tan\beta=50$ and
$10$, dot-dashed line represents the case of switching off
$C_{Q_i}$ contributions.} \label{fig4}
\end{figure}

\begin{figure}
\epsfxsize=15 cm \centerline{\epsffile{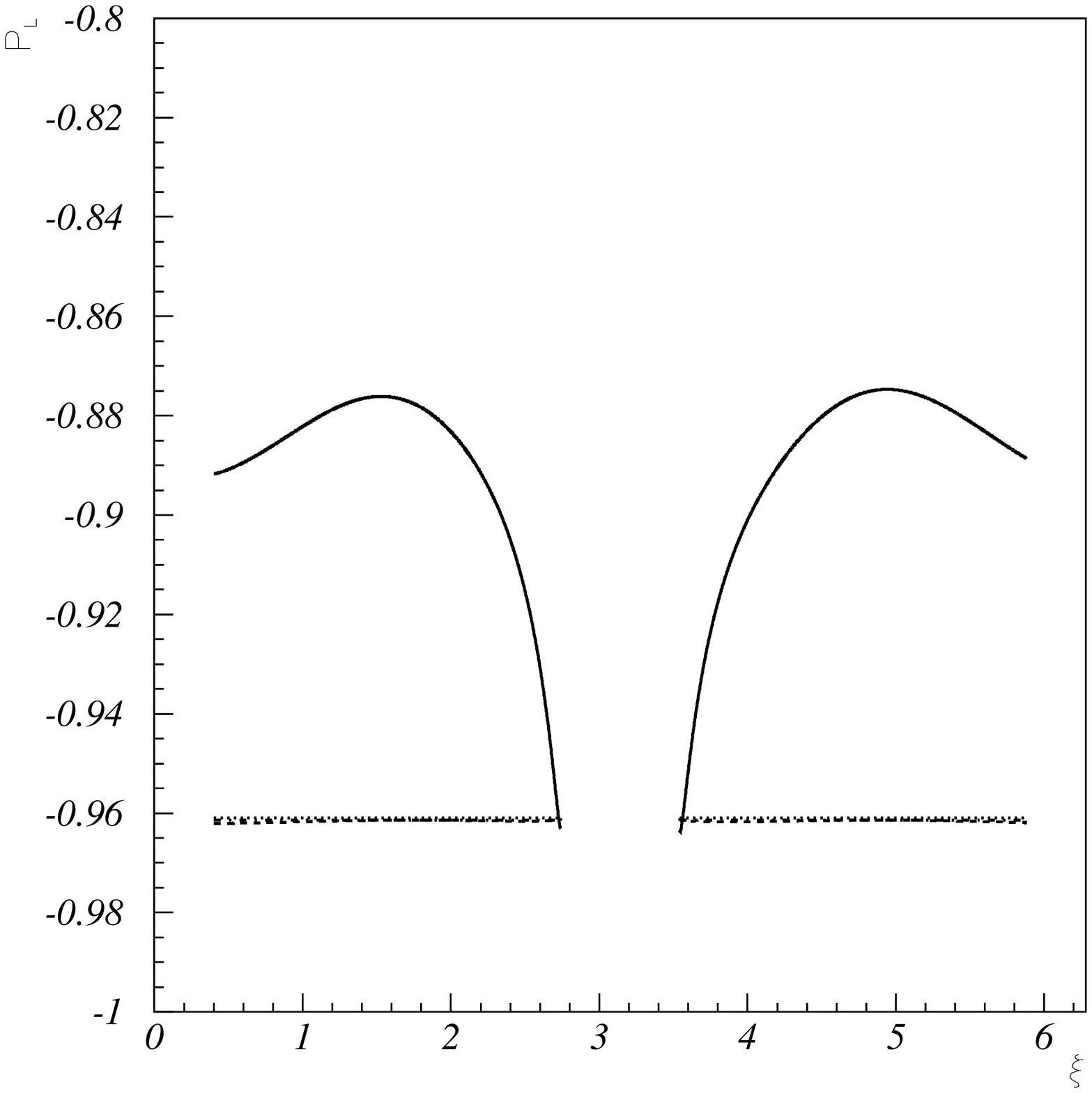}} \caption[]{ $P_L$
as function of $\xi$, other captions are same as Fig. \ref{fig4} (a).}
\end{figure}

\begin{figure}
\epsfxsize=15 cm
\centerline{\epsffile{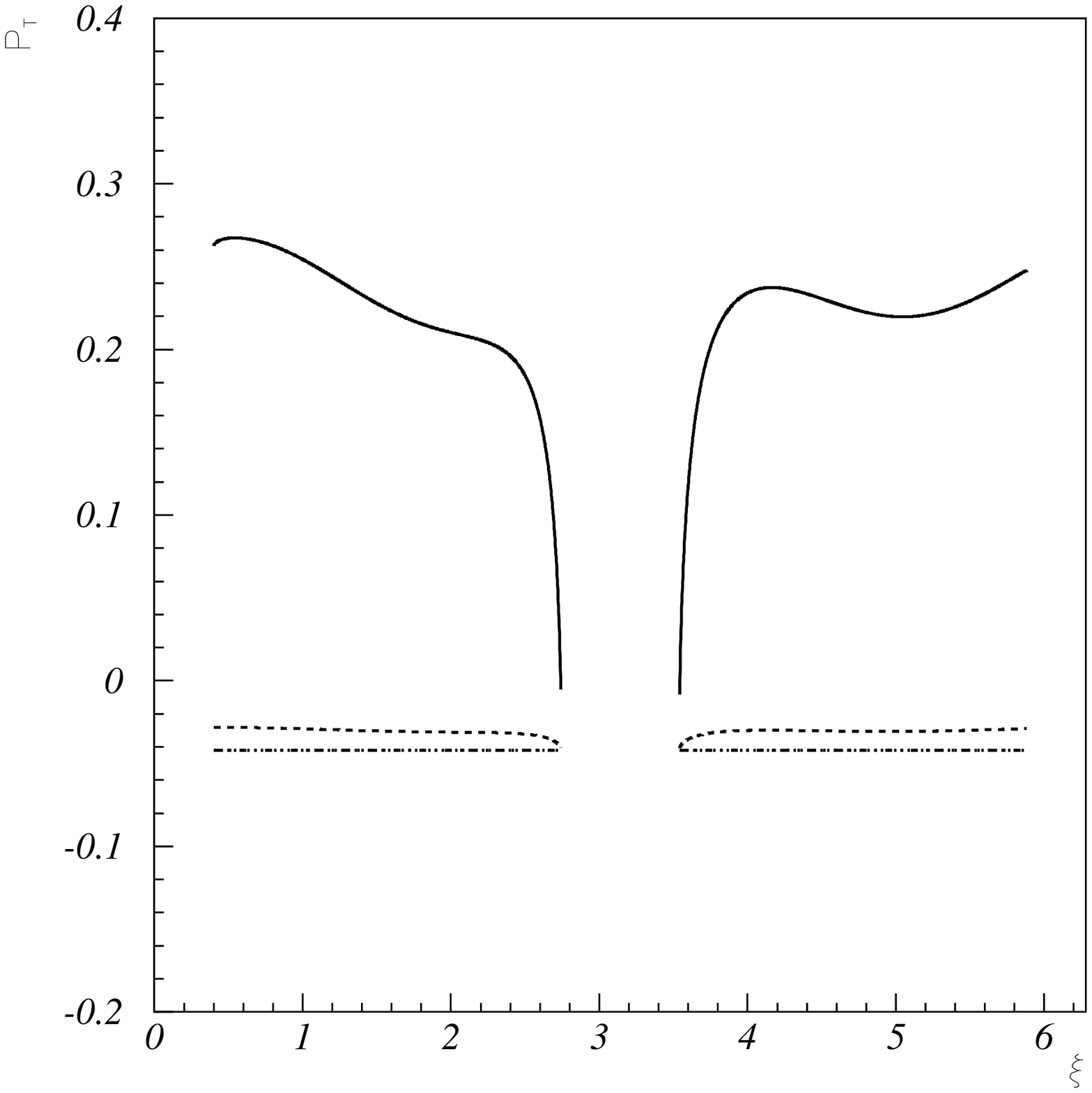}}
\caption[]{
$P_T$ as function of $\xi$,
other captions are same as Fig. \ref{fig4} (a).}
\end{figure}


\end{document}